\documentclass[superscriptaddress,prb,twocolumn,noshowpacs]{revtex4}
\usepackage{graphicx}
\usepackage{amsmath}
\usepackage{bm}

\begin{document}
\title{High-pressure phase diagram, structural transitions, and persistent non-metallicity of BaBiO$_3$: theory and experiment}

\author{Roman Marto\v{n}\'{a}k}
\affiliation{Department of Experimental Physics, Comenius University, 
Mlynsk\'{a} Dolina F2, 842 48 Bratislava, Slovakia}

\author{Davide Ceresoli}
\affiliation{CNR Institute of Molecular Science and Technology (ISTM),
via Golgi 19, 20133 Milan, Italy}

\author{Tomoko Kagayama} \affiliation{KYOKUGEN, Graduate School of
  Engineering Science, Osaka University, Machikaneyamacho 1-3,
  Toyonaka, Osaka 560-8531, Japan}

\author{Yusuke Matsuda} \affiliation{KYOKUGEN, Graduate School of
  Engineering Science, Osaka University, Machikaneyamacho 1-3,
  Toyonaka, Osaka 560-8531, Japan}

\author{Yuh Yamada} \affiliation{Department of Physics, Faculty of
  Science, Niigata University, 8050, Ikarashi 2-no-cho, Nishi-ku,
  Niigata 950-2181, Japan}

\author{Erio Tosatti} \affiliation{International School for Advanced
  Studies (SISSA) and CNR-IOM Democritos, Via Bonomea 265, I-34136
  Trieste, Italy}

\affiliation{The Abdus Salam International Centre for Theoretical
  Physics (ICTP), Strada Costiera 11, I-34151 Trieste, Italy}

\date{\today} 

\begin{abstract} 

BaBiO$_3$ is a mixed-valence perovskite which escapes the metallic
state through a Bi valence (and Bi-O bond) disproportionation or CDW
distortion, resulting in a semiconductor with a gap of 0.8 eV at zero
pressure. The evolution of structural and electronic properties at
high pressure is, however, largely unknown. Pressure, one might have
hoped, could reduce the disproportionation, making the two Bi ions
equivalent and bringing the system closer to metallicity or even to
superconductivity, such as is attained at ambient pressure upon metal
doping. We address the high-pressure phase diagram of pristine
BaBiO$_3$ by ab initio DFT calculations based on GGA and hybrid
functionals in combination with crystal structure prediction methods
based on evolutionary algorithms, molecular dynamics and metadynamics.
The calculated phase diagram from 0 to 50 GPa indicates that pristine
BaBiO$_3$ resists metallization under pressure, undergoing instead at
room temperature structural phase transitions from monoclinic
\textit{I2/m} to nearly tetragonal \textit{P-1} at 7 GPa, possibly to
monoclinic \textit{C2/m} at 27 GPa, and to triclinic \textit{P1} at 43
GPa.  Remarkably, all these phases sustain and in fact increase the
inequivalence of two Bi neighboring sites and of their Bi-O bonds and,
in all cases except semimetallic \textit{C2/m}, the associated
insulating character.  We then present high-pressure resistivity data
which generally corroborate these results, and show that the
insulating character persists at least up to 80 GPa, suggesting that
the \textit{C2/m} phase is probably an artifact of the small
computational cell.

\end{abstract}

\maketitle

\section{Introduction}
BaBiO$_3$ is a narrow gap insulating perovskite of considerable
importance on several accounts.  The first is that the insulating
state is accompanied, and in fact caused, by a relatively small
displacive structural dimerizing CDW-like lattice
distortion,~\cite{CoxSleight1976} also amounting to a spontaneous
static mixed valence disproportionation of two otherwise equivalent Bi
cations from nominal valency 4+ to 3+/5+ \cite{Rice_Sneddon1981} (see
also Fig.1 in Ref.\cite{Thonhauser2006}).  Another is that the
insulating and distorted state reverts to an undistorted metal upon
hole doping in alloys such as Ba$_{1-x}$K$_x$BiO$_3$ yielding an
important class of superconductors, with T$_c$ as high as
30~K\cite{Cava1988, Mattheiss1988}. Superconductivity in these systems
has been discussed as a classic example of what came to be called a
bipolaron state, whereby quantum fluctuations turn the static electron
pairing around the fixed Bi(3+) ion sublattice into a dynamical
state. The quantum mechanically delocalized electron pairs can then
make all Bi sites dynamically equivalent in the superconducting
state\cite{Rice_Sneddon1981}.

The structural behaviour of BaBiO$_3$ is non-trivial already at zero
pressure where it undergoes three phase transitions with
temperature. The low temperature structure with space group
\textit{P2$_1$/n} was first resolved by Kennedy et
al.\cite{Kennedy2006} by neutron diffraction and is stable below 132
K, where it transforms to the \textit{I2/m} structure which is the
room temperature form studied by Cox and
Sleight\cite{CoxSleight1976}. This form is stable until 430 K where it
further transforms into a rhombohedral \textit{R-3} structure. Finally
at 820 K BaBiO$_3$ transforms to a cubic structure with space group
\textit{Fm-3m}.  The bond disproportionation persists all along these
phases.  The main difference between them consists of different
rotation patterns of the BiO$_6$ octahedra\citep{Kennedy2006},
rotations which must therefore involve small energy scales of the
order of the thermal energy $\sim$ 0.01 eV. Moreover, external
perturbations including e.g., friction and strain could easily provoke
transitions between them. For a recent review of structural and
electronic aspects see Ref.\cite{Sleight_review2015}.

BaBiO$_3$ has been the subject of numerous theoretical
studies,~\cite{Mattheiss1983,Mattheiss1988,KUNC1991,Thonhauser2006}
initially underlining the difficulty to reproduce within standard
density functional theory (DFT) the dimerized insulating state at zero
pressure, instead of an undimerized metal.  Subsequent work showed
that the problem is to some extent cured by appropriate choices of the
exchange-correlation potential~\cite{Franchini2010} or by LDA+U
underlining the delicate role of correlations in the zero-pressure
state.~\cite{Korotin2012, Ceresoli_unpubl}.  None of that work,
however, considered a prediction of the high-pressure phase diagram
until the recent unpublished work of Smolyanyuk et
al~\cite{Smolyanyuk2017} which does not contain experimental data but
with whose theoretical results we make good reciprocal contact.

Experimentally, Akhtar et al\cite{AKHTAR1994} provided an equation of
state of pristine BaBiO$_3$ up to 10 GPa. Sugiura and
Yamadaya\cite{Sugiura1986} went up to 20 GPa and found X-ray
diffraction and Raman evidence of a structural transition at 8-10 GPa
to an unidentified tetragonal phase. Other work
\cite{Liu2003,Imai2005} suggested a decrease of electronic gap for
pressures up to 9 GPa, thus still within the initial \textit{I2/m}
phase.  What phase or phases will appear above 10 GPa, whether they
would give up the disproportionation and metallize, or even
superconduct, is still unknown.

We here explore these questions, both theoretically by ab initio
calculations and experimentally by high-pressure resistivity
measurements. Calculations including total-energy based evolutionary
structure search\cite{Oganov2006} as well as molecular dynamics and metadynamics\cite{metadynamics,bpr2003,natmat2006}
simulations, indicate at room temperature a first structural phase
transition from \textit{I2/m} to a nearly tetragonal \textit{P-1}
symmetry around 7 GPa, whose surprising offshoot is that
disproportionation of the two Bi cations as well as insulating
character are preserved and actually enhanced by pressure, rather than
eliminated.  Upon further increase of pressure two rather unusual
structures with space groups \textit{C2/m} and \textit{P1} are found,
both of them disproportionated even if only the second frankly
insulating.  Experimental resistivity although unable to address the
structure, shows near 10 GPa a phase transition from an insulating
phase to another, with no metallic phases and a gap increasing at
least until 40 GPa when the gap levels off.

The paper is organized as follows. In Section II we provide some
details about the computational methodology used. In Section III we
describe the experimental measurement of resistivity at high
pressure. In Section IV we show the results of our evolutionary
structural search and MD and metadynamics simulations, discuss the 
structural and electronic properties of the new phases and compare to 
experimental data. In Section V we draw our conclusions.

  \begin{figure}[ht]
    \centering
    (a) \includegraphics[width=0.2\textwidth]{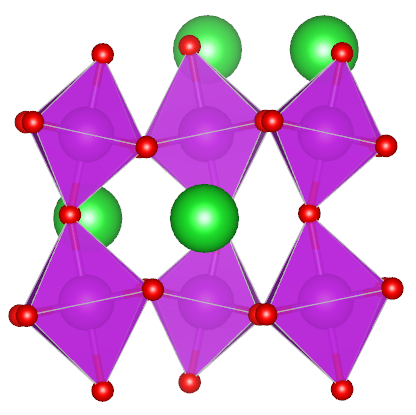}
     \includegraphics[width=0.2\textwidth]{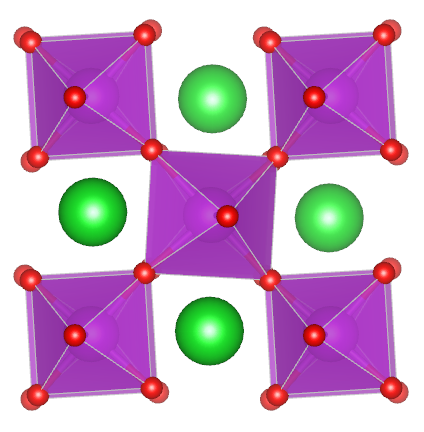}\\    
    (b) \includegraphics[width=0.2\textwidth]{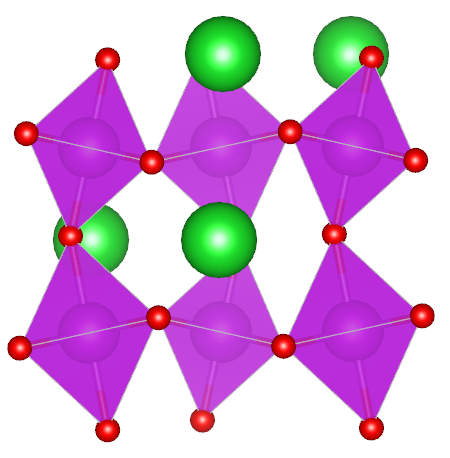}
    \includegraphics[width=0.2\textwidth]{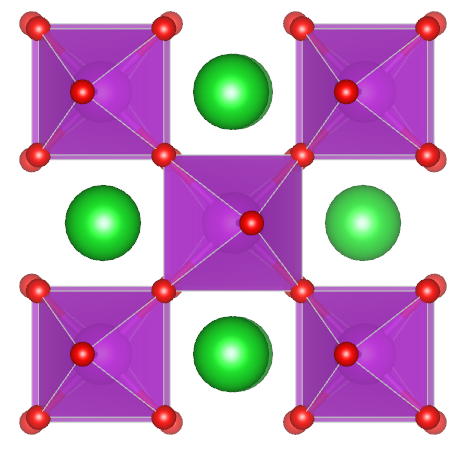}\\
    (c) \includegraphics[width=0.2\textwidth]{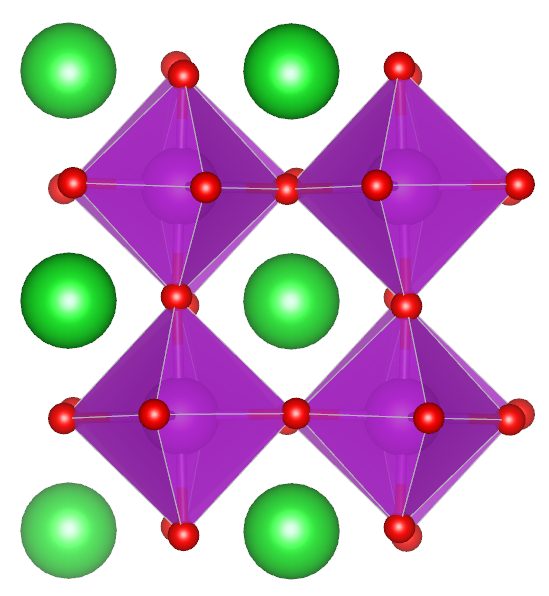}
    \includegraphics[width=0.2\textwidth]{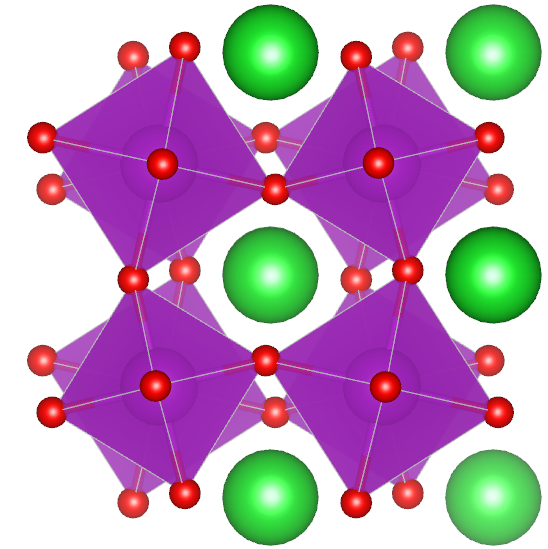}\\
    (d) \includegraphics[width=0.2\textwidth]{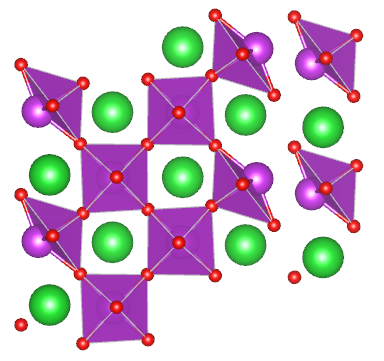}
    \includegraphics[width=0.2\textwidth]{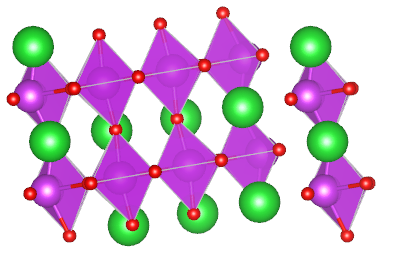}\\
    (e) \includegraphics[width=0.2\textwidth]{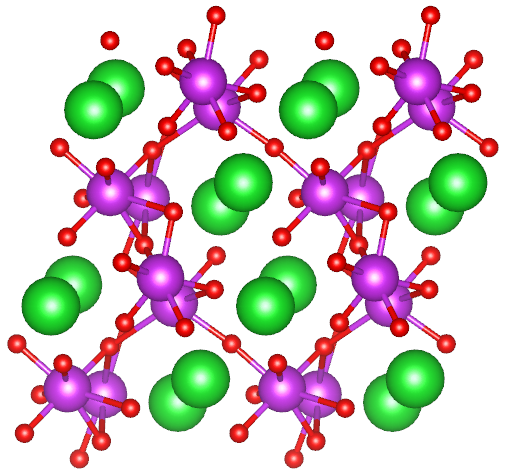}
    \includegraphics[width=0.2\textwidth]{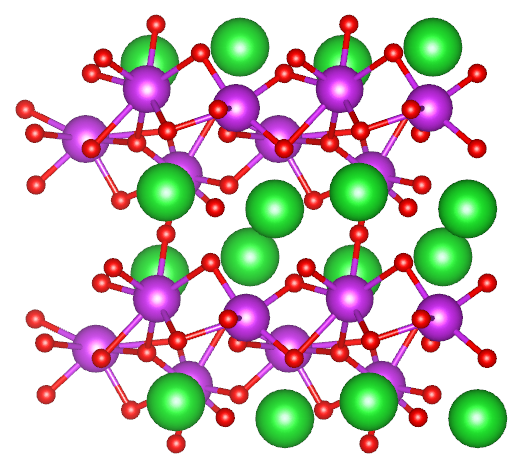}
    \caption{\textit{P2$_1$/n} at 0 GPa (a), \textit{I2/m} at 0 GPa (b),
    \textit{P-1} at 20 GPa (c), \textit{C2/m} at 30 GPa (d) and
      \textit{P1} (e) at 50 GPa, respectively. Note that \textit{C2/m} 
      can be obtained from ideal perovskite by creating a stacking fault 
      and \textit{P1} is disordered.}
    \label{fig:structures}
  \end{figure}

\section{Computational Methods}
\label{sec:computational}

Crystal structure prediction in this system represents a computational
challenge. One reason is the difficulty of standard LDA or GGA
functionals to properly describe the energetics of the
system.~\cite{Mattheiss1983,Mattheiss1988,KUNC1991,Thonhauser2006}
While this appears to be cured by using the hybrid
functional,~\cite{Franchini2010} it is not easy to perform
evolutionary search using the latter because of its high computational
cost. Moreover, the fact that already at $p=0$ the system between zero
and room temperature undergoes a temperature-induced structural
transition points to the importance of entropic effects, neglected in
standard prediction schemes based on e.g. evolutionary algorithms.
For this reason we complement our study with molecular dynamics
simulations, where entropy is included. In order to study structural
transformations including kinetic effects we also apply the
metadynamics\cite{metadynamics}-based approach to
structural phase transitions simulation\cite{bpr2003,natmat2006} which
is able to cross barriers and dynamically follow structural
transformations.

We performed the evolutionary crystal structure search by using the
Xtalopt code\cite{xtalopt} in combination with the VASP
package\cite{vasp,vasp_paw}. All calculations employed PAW
pseudopotentials with 10 valence electrons (5s$^2$5p$^6$6s$^2$) for
Ba, 15 electrons (5d$^{10}$6s$^2$6p$^3$) for Bi and 6 electrons
(2s$^2$2p$^4$) for O. During the search the exchange-correlation
energy was calculated within the PBE \cite{PBE} functional and an
energy cutoff of 520 eV. The structural optimization was performed in
several steps using tighter convergence criteria and denser
Monkhorst-Pack\cite{mpgrid} grids in later steps.  To begin with,
below 10 GPa the stability and structure of both \textit{P2$_1$/n} and
\textit{I2/m} phases were checked and optimized.  The subsequent
structural search was carried out at pressures of 20, 30 and 40 GPa
with a supercell containing four formula units (20 atoms). At both 20
GPa and 30 GPa the search generated more than 2000 structures. At 40
GPa we first generated more than 1000 structures. Afterwards we
selected eight best structures and doubled the cell in one direction,
thus creating cells with 40 atoms. These eight structures were
subsequently used as seeds for an additional search at the same
pressure which generated more than 1500 structures. The 40 atom
supercell search, however, yielded no better structure than the 20
atoms cell search. The zero
temperature structural search results should therefore be considered
reliable within the usual limitations of limited number of structures
generated and limited number of atoms in unit cell.

The most favorable structures generated by the search were
subsequently refined by high accuracy electronic structure
calculations. Structures, enthalpies and band structures with careful
determinations of band gaps were obtained by using the HSE06 hybrid
functional\citep{krukau,hse06}, shown by Franchini et
al. \cite{Franchini2010} to correctly reproduce structural parameters,
bond disproportionation and band gaps.

\section{Experimental - High Pressure Resistivity } 

Single crystals of BaBiO$_3$ were grown by cooling a melt of
polycrystalline specimen in an alumina crucible from 900 to 600
$^{\circ}$C at a rate of 2 $^{\circ}$C/h in a higher oxygen
atmosphere.  The pressure was generated by the diamond anvil cell
technique with a pair of diamond anvils of 150 microns culet. A gasket
was made by a holed (70 micron in diameter) rhenium sheet subsequently
covered with a mixture of cubic-boron-nitride powder and epoxy resin
on a surface which plays both roles of a pressure transmitting media
and electrical insulator to electrodes.  The electrical resistance was
measured with an electrometer using two platinum foil strip
electrodes.

\begin{figure}[ht]
    \centering
    \includegraphics[width=0.45\textwidth]{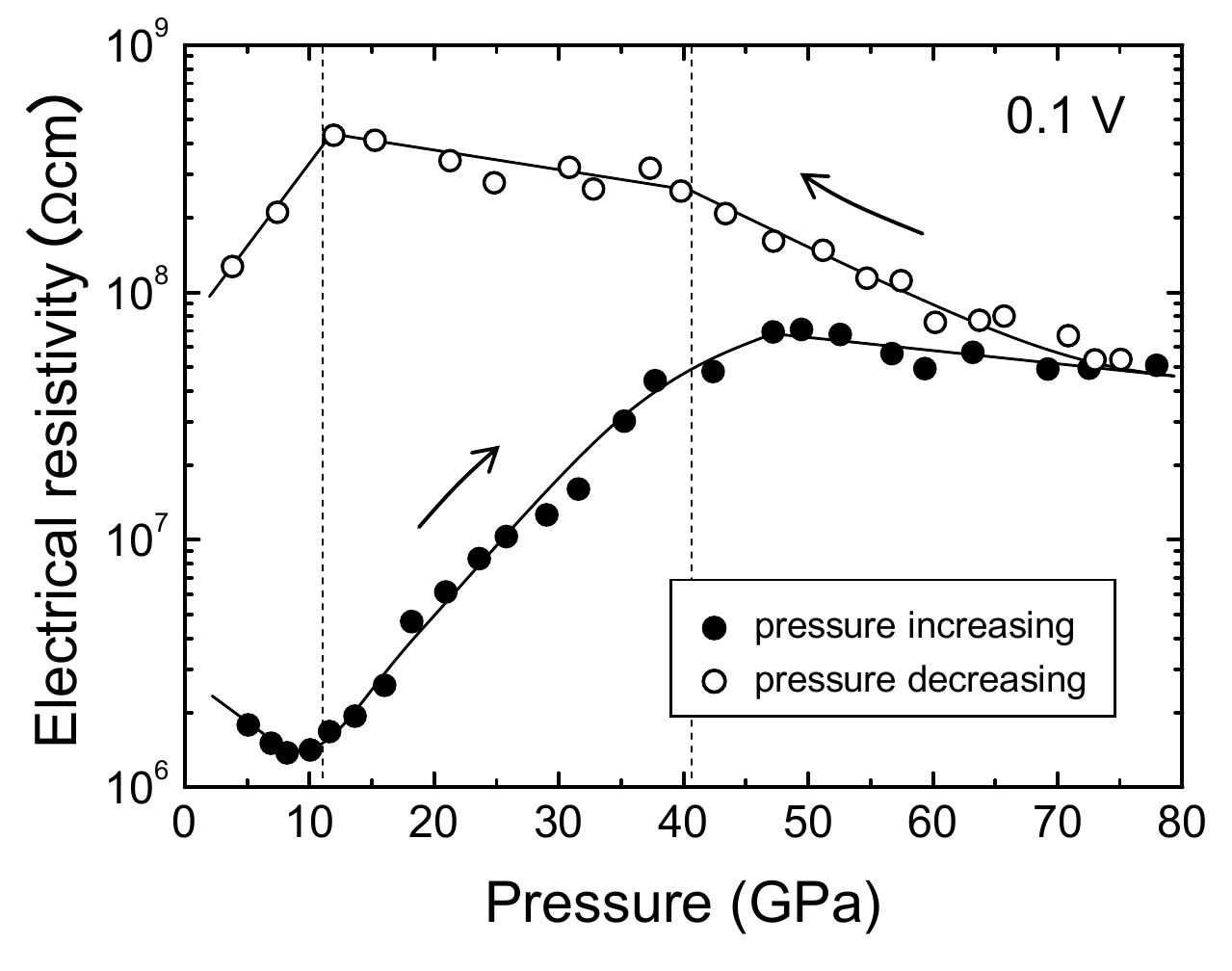}
    \caption{Experimental data for resistivity of BaBiO$_3$ under pressure.}
    \label{fig:resistivity}
  \end{figure}

The pressure was determined by a fluorescence method with a ruby chip
mounted on the sample as a standard material.  The pressure dependence
of the resistivity in the interval from 0 to 80 GPa is shown in
Fig.\ref{fig:resistivity}.

The resistivity initially decreases, in agreement with earlier data by
Imai\cite{Imai2005} and at 8 GPa reaches about 80\% of the value at 5
GPa. Beyond 8 GPa it, however, starts to grow and the ratio of
resistivity at 40 and 10 GPa is about 35 (Fig.\ref{fig:resistivity}).
Assuming a simple exponential dependence of resistivity on the gap
these numbers at room temperature imply an initial decrease of the gap
between 5 and 8 GPa by 6 meV followed by increase by 92 meV between 10
and 40 GPa. The latter increase corresponds to the slope of about 30
meV/10 GPa.  A large hysteretic behaviour is clearly seen upon
pressure decrease suggesting that pressurization induces a major
change in the system, possibly a reconstructive transition resulting
in a barrier preventing the high-pressure state from returning to the
original state.

\section{Theoretical Results} 

The structural search performed at pressures of 20, 30 and 40 GPa
provided several new low enthalpy structures, nearly monoclinic
\textit{C2/c} and nearly tetragonal \textit{P-1}, monoclinic
\textit{C2/m} and triclinic \textit{P1}.  In order to determine the
phase stability at finite temperature one should compare the Gibbs
free energies of the respective phases. An accurate calculation of
free energies in this case is, however, difficult, for the reasons
mentioned in Sec.II, especially at low pressures where the PBE
functional might not be reliable, and the room temperature entropic
effects are larger.  We therefore limit ourselves to comparison of
enthalpies shown in Fig.\ref{fig:enthalpies} and we will comment about
possible thermal effects where appropriate.

\subsection{Crystal structure}

\subsubsection{Pseudotetragonal P-1 phase}

Evolutionary search at 20 GPa provided several structures with low
\textit{P-1} symmetry and very close enthalpies. Structurally they are
still related to disproportionated perovskite and differ by very tiny
distortions of the octahedra. The lowest enthalpy structure among them
is close to monoclinic symmetry \textit{C2/c} followed by several
structures with enthalpy higher only by up to 2.3 meV per f.u. (0.46
meV per atom) which are very close to tetragonal symmetry and within a
tolerance of 0.06 A can be identified with space group \textit{I4/m}.
These minute enthalpy differences are not likely to survive at room
temperature where they are likely to be washed out by thermal
fluctuations. The common feature of all these structures is that they
represent small variations of tetragonal perovskite structure similar
to what SrTiO$_3$ adopts below 110 K~\cite{Unoki1967} where octahedra
are rotated around $z$-axis in an antiferrodistortive manner (space
group \textit{I4/mcm}) -- the additional feature here being the bond
disproportionation.  Since in Ref.\cite{Sugiura1986} a transition to a
tetragonal structure was reported between 8-10 GPa we consider the
pseudotetragonal P-1 structures (see Fig.\ref{fig:structures}) more
representative of the room temperature structure above 10
GPa. Therefore we limited ourselves to pick one of these structures
and use it in subsequent calculations\footnote{Cif files with
  structural data of the new structures are in Supplementary
  material.}.

\begin{figure}[ht]
    \centering
    \includegraphics[width=0.45\textwidth]{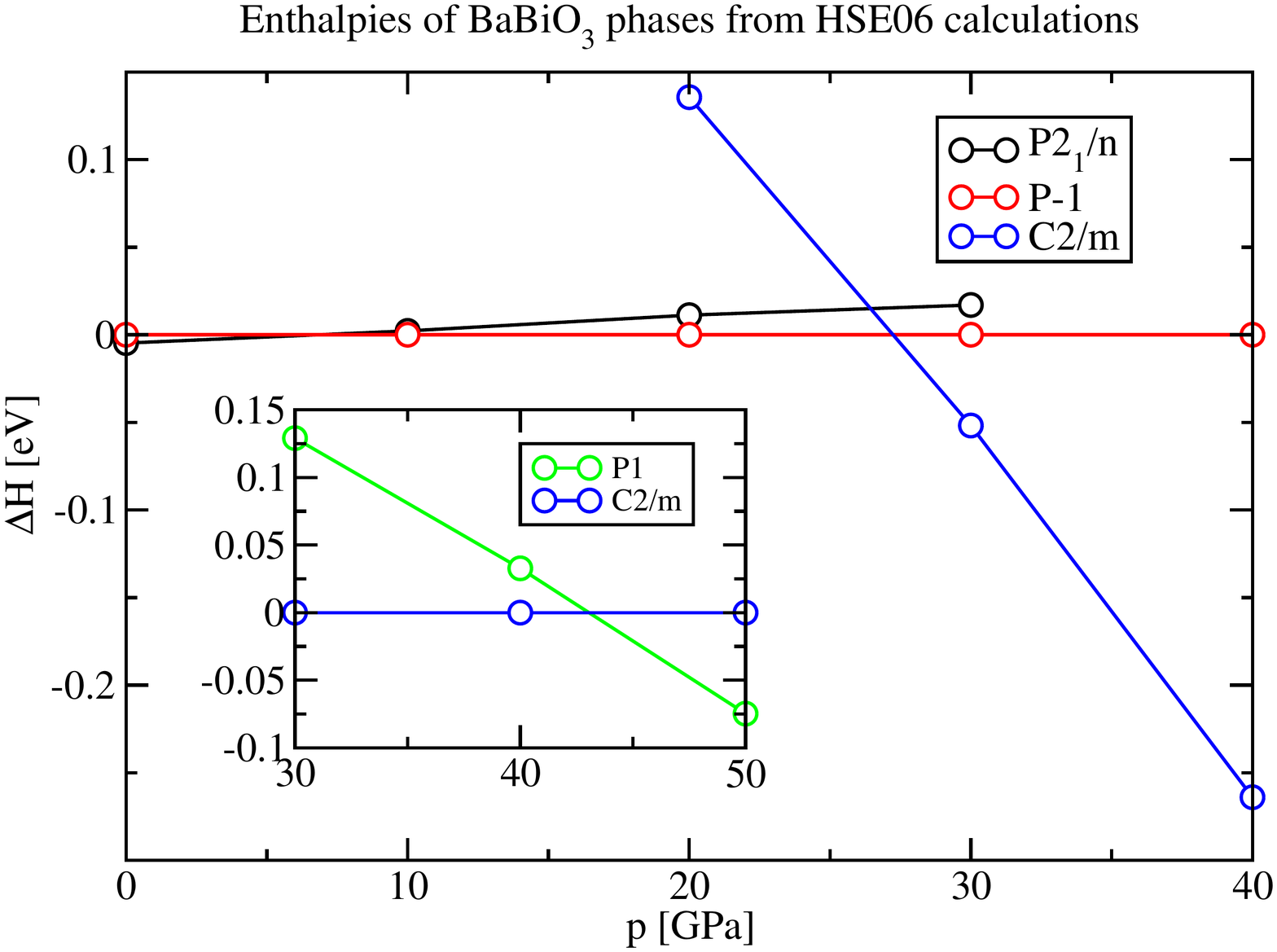}
    \caption{Enthalpies (per formula unit) of various BaBiO$_3$ phases
      as function of pressure. In the main figure enthalpies are
      relative to the \textit{P-1} phase and the inset shows the
      enthalpy of the \textit{P1} phase relative to the \textit{C2/m}
      phase.}
    \label{fig:enthalpies}
  \end{figure}

According to the enthalpy plot (Fig.\ref{fig:enthalpies}) the
\textit{P-1} phase replaces the \textit{P2$_{1}$/n} phase at $p = 7$
GPa -- note that \textit{P-1} has a slightly smaller volume (by 0.2 \%
at 10 GPa) compared to the \textit{P2$_{1}$/n} phase. The
disproportionation pattern of Bi-O bonds in \textit{P-1} is identical
to that of the low-pressure \textit{P2$_{1}$/n} and \textit{I2/m}
structures where expanded and collapsed octahedra are arranged in an
fcc-like arrangement (similar to rocksalt). The octahedra are rotated
with respect to the ideal perovskite structure around the $z$ axis, by
an average rotation angle of 13.15$^\circ$.
(Fig.\ref{fig:structures}).

\begin{figure}[ht]
    \centering
    \includegraphics[width=0.45\textwidth]{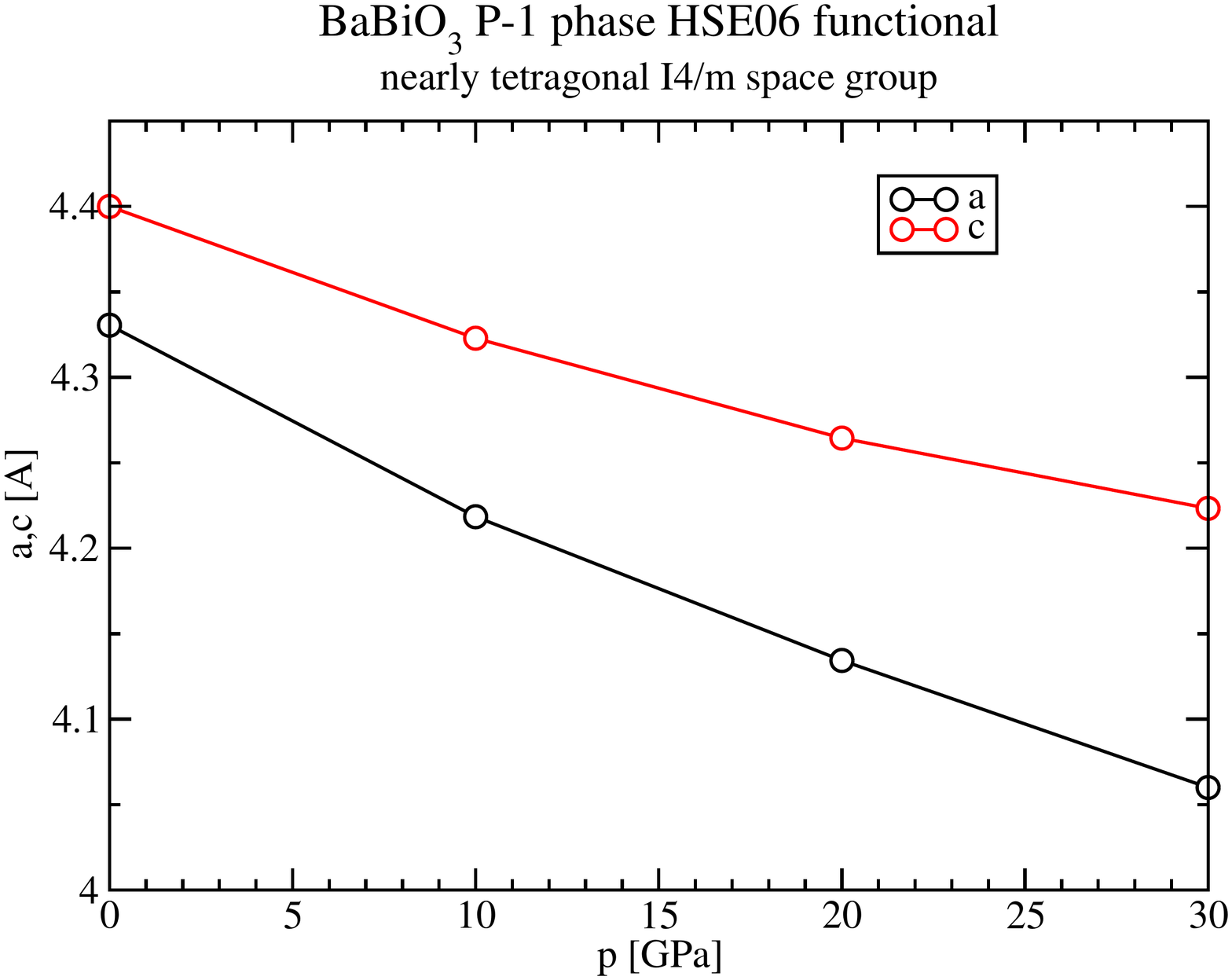}
    \caption{Pressure dependence of the pseudotetragonal lattice parameters of P-1 structure.}
    \label{fig:lattice_param}
  \end{figure}

Our predicted pseudotetragonal \textit{P-1} phase can be favorably
compared to the available experimental data for the lattice parameters
of the unidentified tetragonal structure discovered by Sugiura and
Yamadaya\cite{Sugiura1986} at 8-10 GPa. In Fig.\ref{fig:lattice_param}
we show the pressure dependence of pseudotetragonal lattice parameters
$a,c$ of our zero-temperature \textit{P-1} phase. With increasing
pressure the $c/a$ ratio increases deviating further from 1, a feature
which qualitatively agrees with the experimental data in
Ref.\cite{Sugiura1986}, Fig.1.

Since the zero temperature \textit{P2$_1$/n} phase is replaced by
\textit{I2/m} at room temperature, it is possible that \textit{P-1}
might also undergo some kind of transition between zero and room
temperature.  In order to assess the importance of thermal effects we
carried out an ab-initio variable-cell NPT molecular dynamics (MD)
simulation at $p=20$ GPa and $T=300$ K employing a Parrinello-Rahman
barostat\cite{pr} and a Langevin thermostat.  The MD simulation was
started from the ideal perovskite structure in cubic $2\times2\times2$
supercell (40 atoms) and dynamically simulated for 15000 steps
representing a 30 ps time interval.  Since the barriers between
structures differing by octahedra rotation pattern in such small
system are likely to be small the procedure may be expected to provide
a realistic result for the room temperature structure.

\begin{figure}[ht]
    \centering
    \includegraphics[width=0.45\textwidth]{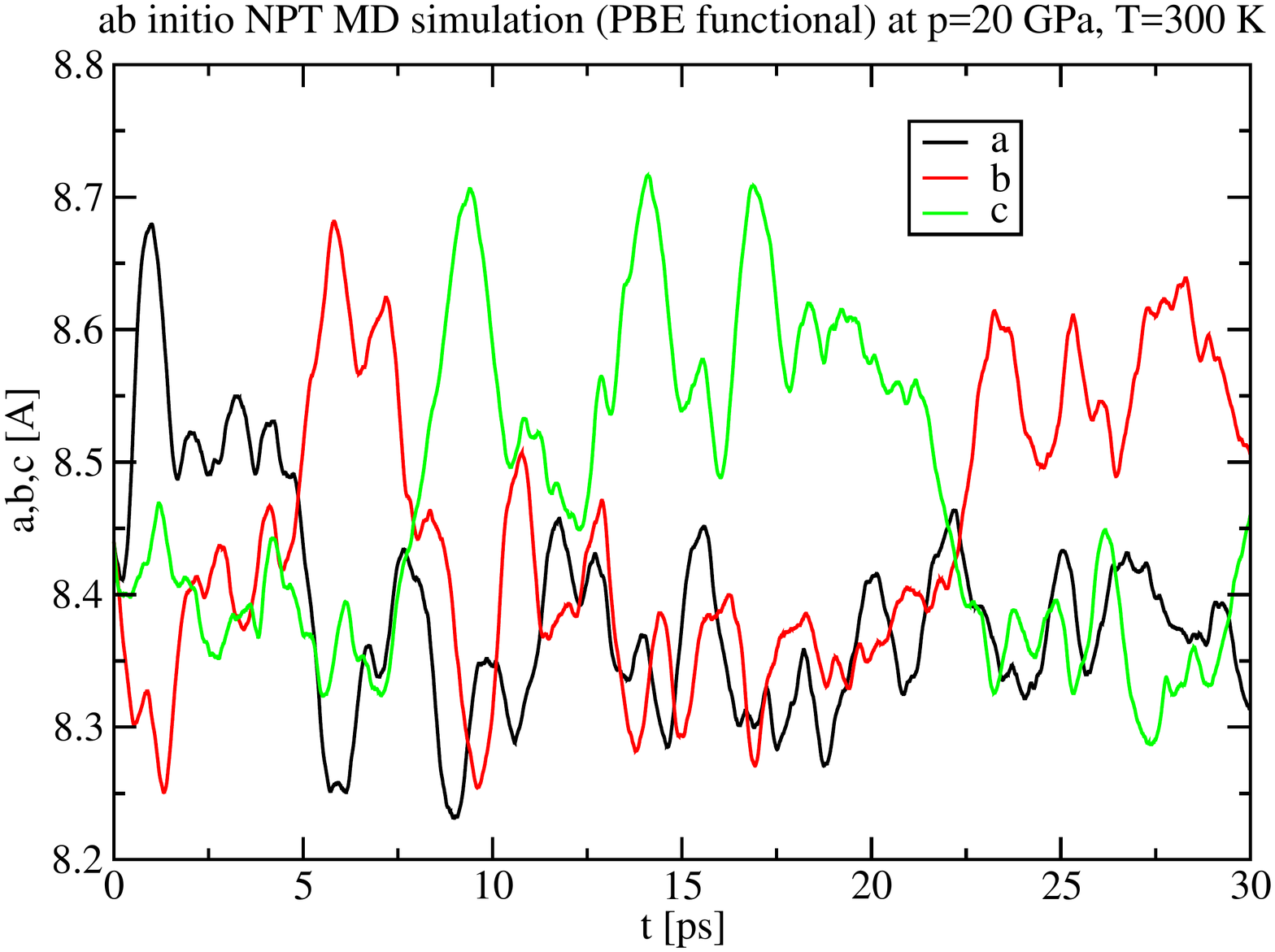}
    \caption{Time evolution of supercell parameters during MD run at
      $p=20$ GPa and $T=300$ K.}
    \label{fig:md}
  \end{figure}

The time evolution of cell parameters from this run is shown in
Fig.\ref{fig:md}.  It is seen that the lengths of the 3 supercell
edges $a,b,c$, initially equal in the ideal perovskite structure,
quickly split, into one longer and 2 shorter cell edges, the latter
two equal on average.  This transformation is related to the rotation
of octahedra which takes place precisely in tetragonal SrTiO$_3$-like
manner, similar to the pseudotetragonal \textit{P-1} structure. This
gives high confidence that the SrTiO$_3$-like octahedra rotation
pattern is a robust feature and the structure observed in the MD run
indeed represents the stable minimum of the Gibbs free energy at room
temperature and $p=20$ GPa. While a much longer run would be necessary
to accurately determine average values of the cell lengths and angles,
the latter are seen to fluctuate around 90 degrees, a fact which
together with the splitting of the cell lengths suggests that the
\textit{P-1} structure appears tetragonal at room temperature (note
that the tetragonal axis switches among various directions during the
run because of the small size of the system). The $c/a$ ratio from MD
at 20 GPa is estimated to be about 1.025 which is smaller than the zero
temperature value of 1.031 (Fig.\ref{fig:lattice_param}) and very
close to the experimental value of 1.024 from Ref.\cite{Sugiura1986}
(Fig.1).

In order to cross-check our prediction from the evolutionary algorithm
we annealed the final structure from the NPT MD run at 300 K by
lowering the temperature in 2 steps, first to 150 K and then to 30 K
and performing at each temperature a 16 ps NPT MD run. The final structure
from the 30 K run was structurally relaxed at $p=20$ GPa and resulted
again in a \textit{P-1} structure (very close to tetragonal
\textit{I4/m}) with enthalpy only 2.5 meV/f.u.  above the best
\textit{P-1} found at the same pressure by evolutionary search. This
provides a strong independent confirmation of the validity of the
evolutionary search.

\begin{figure}[ht]
    \centering
    \includegraphics[width=0.45\textwidth]{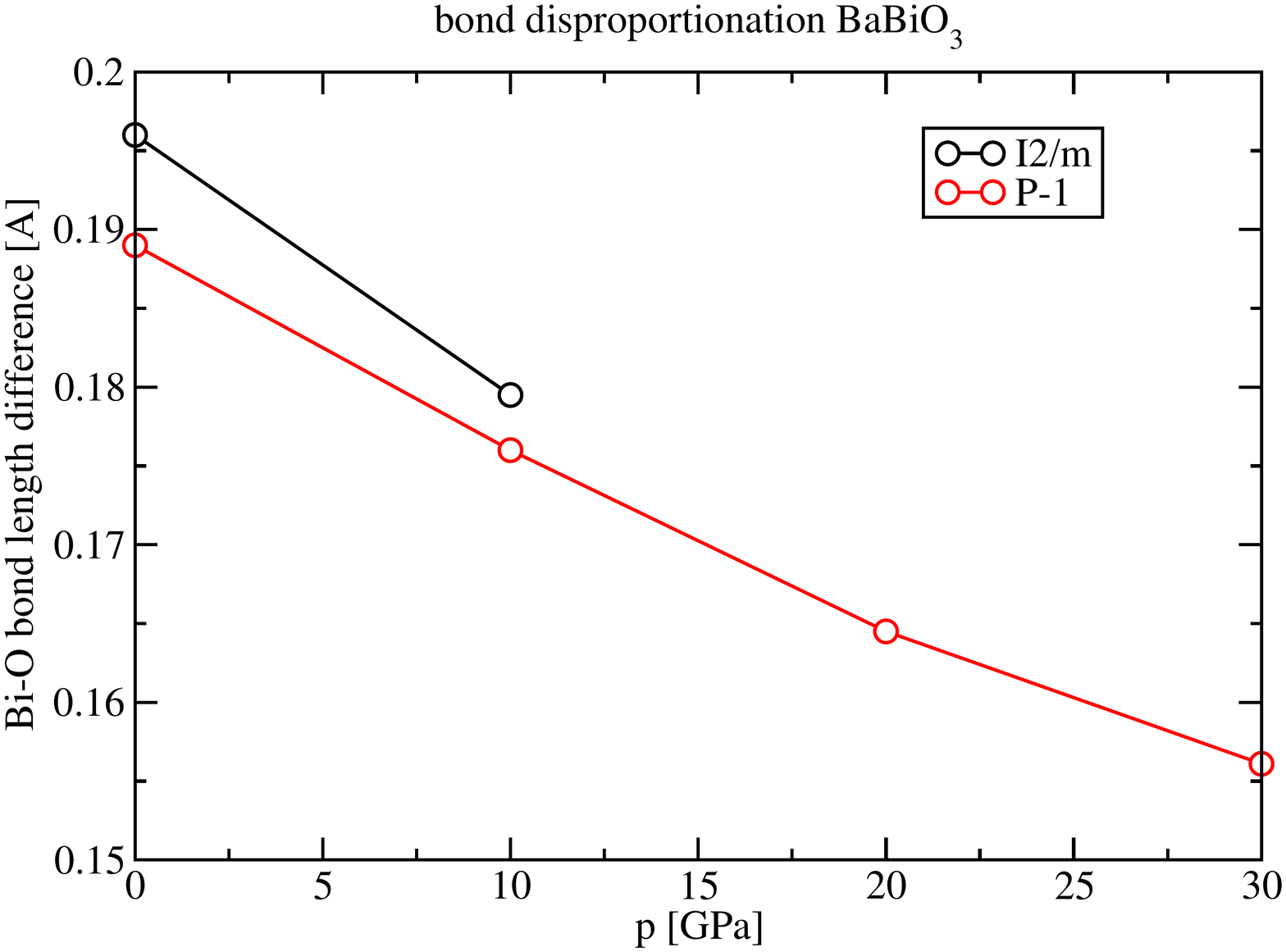}
    \caption{Pressure dependence of the bond disproportionation 
in \textit{I2/m} and \textit{P-1} BaBiO$_3$.}
    \label{fig:disprop}
  \end{figure}

In Fig.\ref{fig:disprop} we show the pressure evolution of the bond
disproportionation defined as difference between the average Bi-O bond
length in expanded and collapsed octahedra, respectively. In both
\textit{I2/m} and \textit{P-1} phases this inbalance is seen to
slightly decrease with increasing pressure.

%
  
  \begin{figure}[ht]
    \centering
     \includegraphics[width=0.2\textwidth]{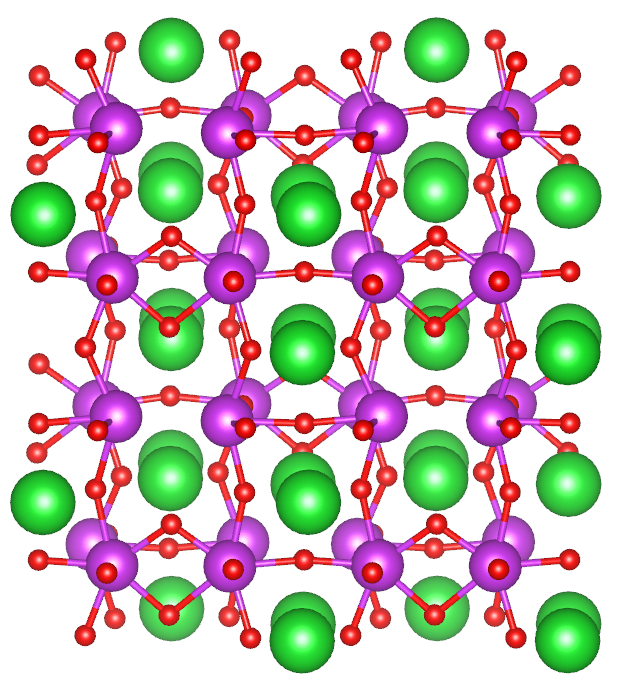}
     \includegraphics[width=0.21\textwidth]{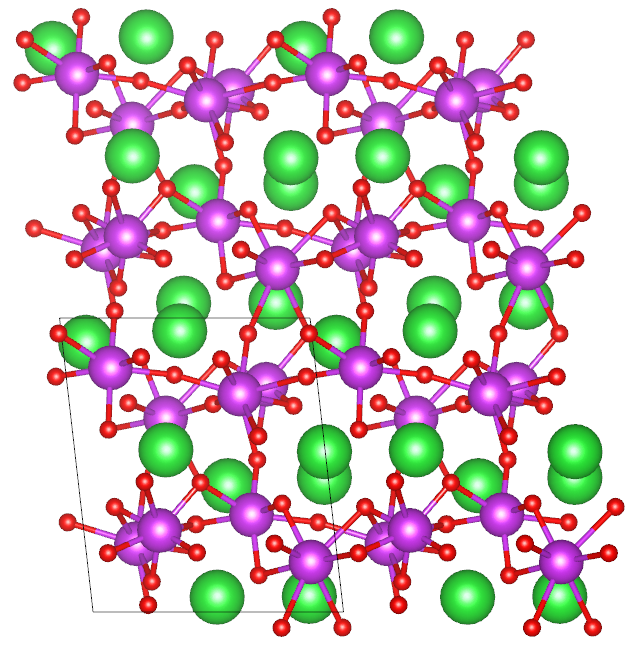}\\
    \caption{Structures found by ab-initio metadynamics starting from the ideal perovskite at $T=300$ K and $p=50$ GPa. First defects are created (left panel) and finally a disordered structure is formed (right panel).}
    \label{fig:mtd}
  \end{figure}

\subsubsection{Monoclinic \textit{C2/m} phase}

A second structure was identified within our evolutionary search at 30 and
40 GPa which, with a space group \textit{C2/m}, can be regarded as a
perovskite structure with a stacking fault (Fig.\ref{fig:structures}).
Its volume at 30 GPa is 4.7 \% lower than that of \textit{P-1} phase.
This phase maintains two inequivalent Bi atoms in a rather unusual
manner, by partly preserving and partly breaking the parent perovskite
structure. This results in different local environments, one close to
perovskite and the other not.  Altogether, here we have a much
stronger inequivalence of the two Bi ions than that realized at lower
pressure by bond disproportionation. According to the enthalpy
calculations it is stable from 27 GPa till 43 GPa
(Fig.\ref{fig:enthalpies}).

\subsubsection{Triclinic \textit{P1} phase}
\label{P1phase}

The next structure identified within the evolutionary search at 40 GPa
has no symmetry and appears totally different from
perovskite. Interestingly, it is nearly layered
(Fig.\ref{fig:structures}). There are still at least two inequivalent
Bi ions, but there are no well defined coordination polyhedra around
them.  The volume of this \textit{P1} phase at 40 GPa is 1.4 \% lower
than that of \textit{C2/m} phase. According to the enthalpy
calculations it is stable with respect to the \textit{C2/m} phase
above 43 GPa (Fig.\ref{fig:enthalpies}).

Unlike the lower pressure \textit{P-1} slightly distorted perovskite,
neither the \textit{C2/m} nor the \textit{P1} phases represent a small
distortion of parent high-symmetry structure.  For this reason they
appear less likely to undergo a structural transformation between zero
and room temperature due to entropic effects.  We note that the
structural disorder in \textit{P1} phase could also be related to a
possible tendency to chemical decomposition at high pressure.

Altogether, both \textit{C2/m} and \textit{P1} structures appear
somewhat dubious. They show either an unusual and composite atomic
arrangement (\textit{C2/m}) or disorder and no symmetries
(\textit{P1}), suggesting that they could be artifacts of the specific
finite-size cell used in evolutionary search, especially if the true
thermodynamically stable crystal structure might have a larger cell,
possibly with higher symmetry (especially in the case of \textit{P1}).
The true ground state might involve a complex arrangement of atoms
only reachable in the search after many more generated structures.
This question must be left for the future since extension of the
present study in either direction is at the moment prohibitively
demanding.

Experimentally, even if there were a complex thermodynamically stable
structure with large unit cell, it might not be easy to reach for
kinetic reasons in high-pressure experiment. It is therefore plausible
that neither \textit{C2/m} and \textit{P1} are seen in
experiment. Instead, \textit{P-1} or some similar form, still related
to perovskite structure, might survive as metastable structure beyond
27 GPa until it reaches the limit of mechanical stability.

In order to investigate what might be happening at 50 GPa and beyond
we applied the metadynamics\cite{metadynamics}-based
approach\cite{bpr2003,natmat2006}.  The simulations were started from
ideal perovskite structure on a supercell with 40 atoms and performed 
at $p=50$ GPa and $T=300$ K. Two typical structures from this simulation 
are shown in Fig.\ref{fig:mtd} and represent a possible experimental outcome 
of pressurizing \textit{P-1}. The structure on the left panel in
Fig.\ref{fig:mtd} represents a heavily distorted perovskite with a
massive Bi disproportionation which further evolves towards disordered
structure with tendency towards creation of layers (right panel in
Fig.\ref{fig:mtd}), in fact similar to the predicted \textit{P1}. The
fact that these structures appear in metadynamics simulation shows
that they are likely much easier to be reached kinetically rather than
the search-predicted \textit{C2/m} and \textit{P1}. We suspect that
the structures observed in the resistivity experiment beyond 50 GPa
might well be of this kind.

\subsection{Electronic structure}

\begin{figure}[ht]
    \centering
    \includegraphics[width=0.45\textwidth]{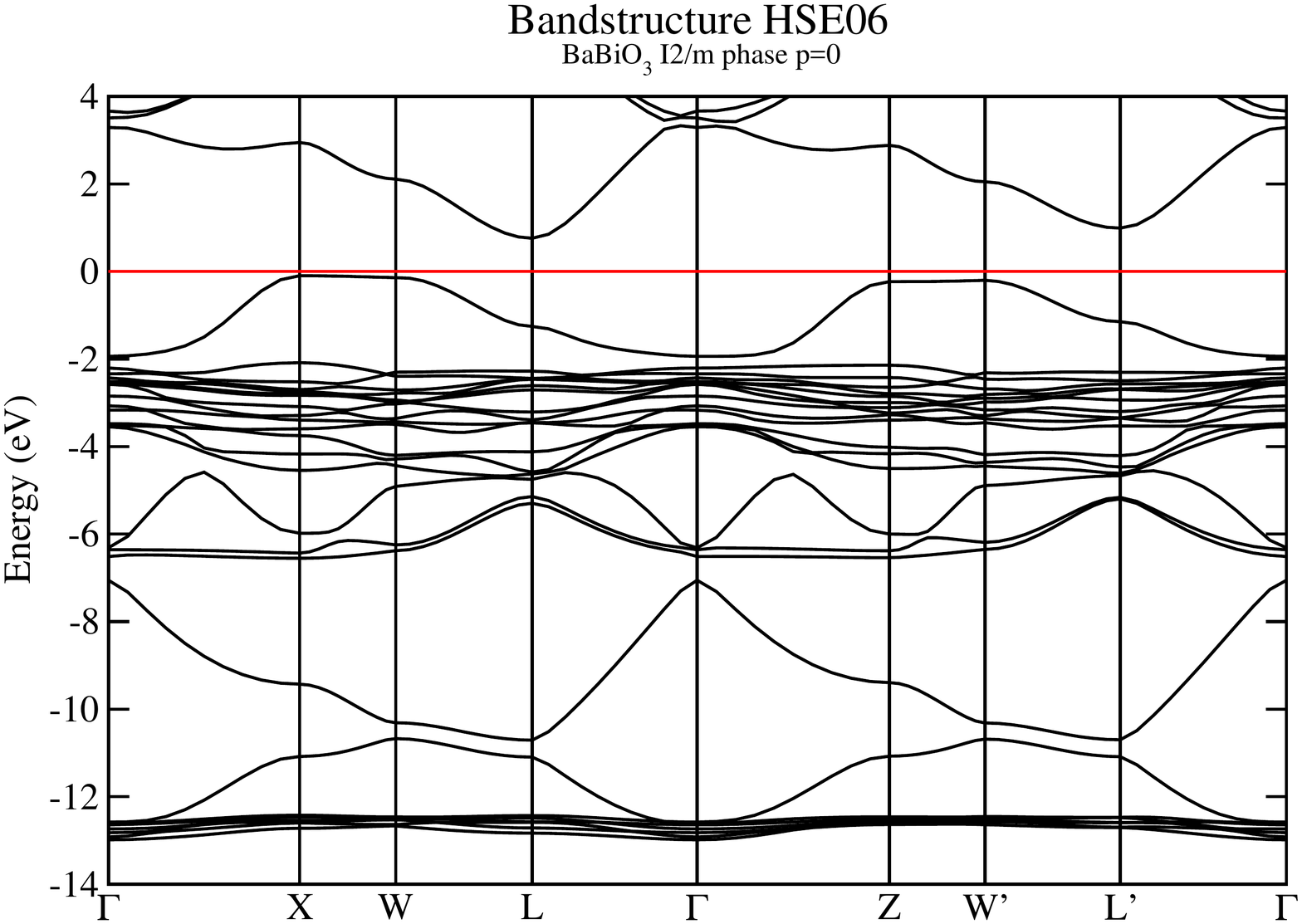}
    \includegraphics[width=0.45\textwidth]{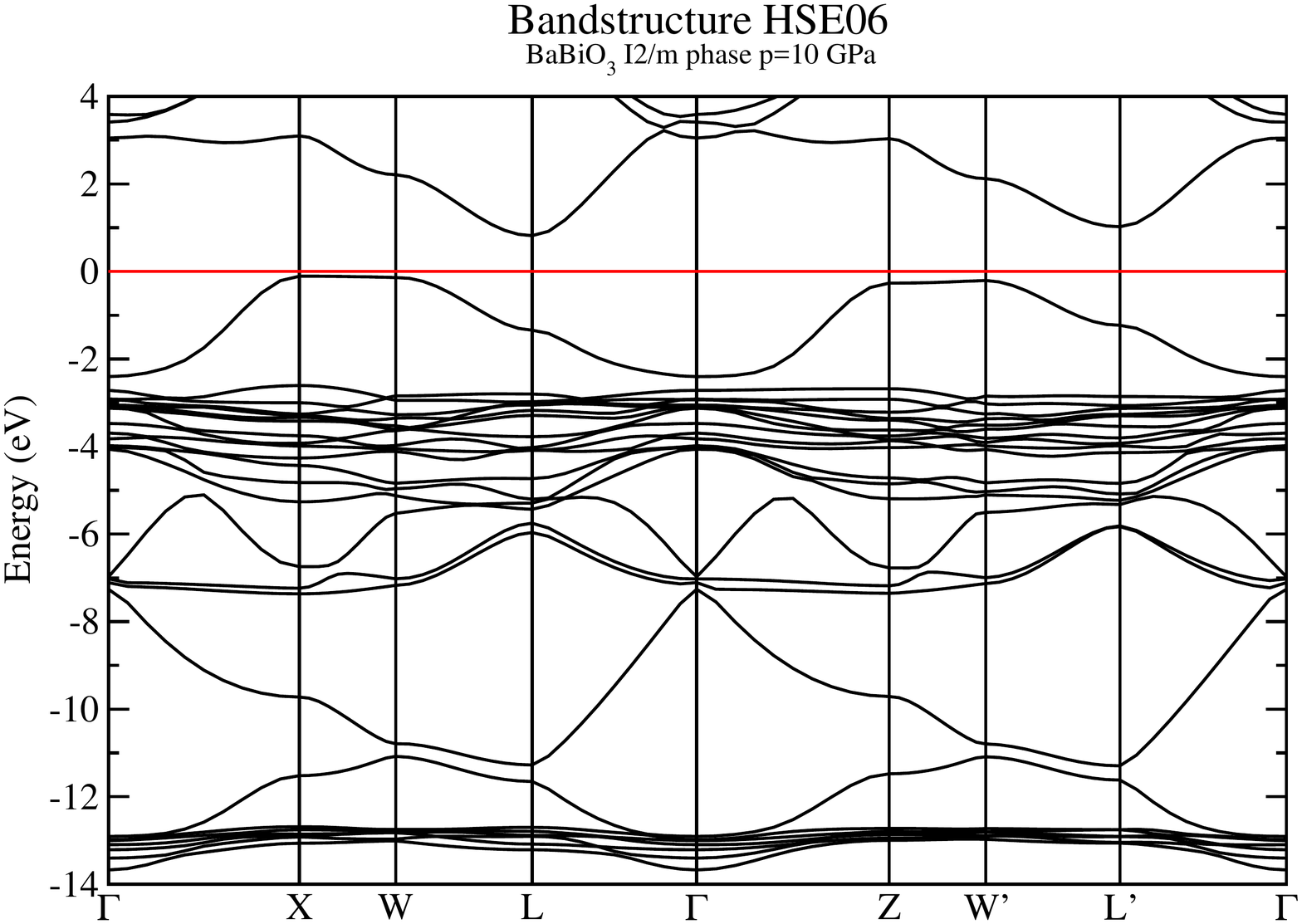}
    \caption{HSE06 electronic structure of \textit{I2/m} BaBiO$_3$ at $P=0$ and $P=10$ GPa.}
    \label{fig:bands_I2/m}
  \end{figure}

\begin{figure}[ht]
    \centering
    \includegraphics[width=0.45\textwidth]{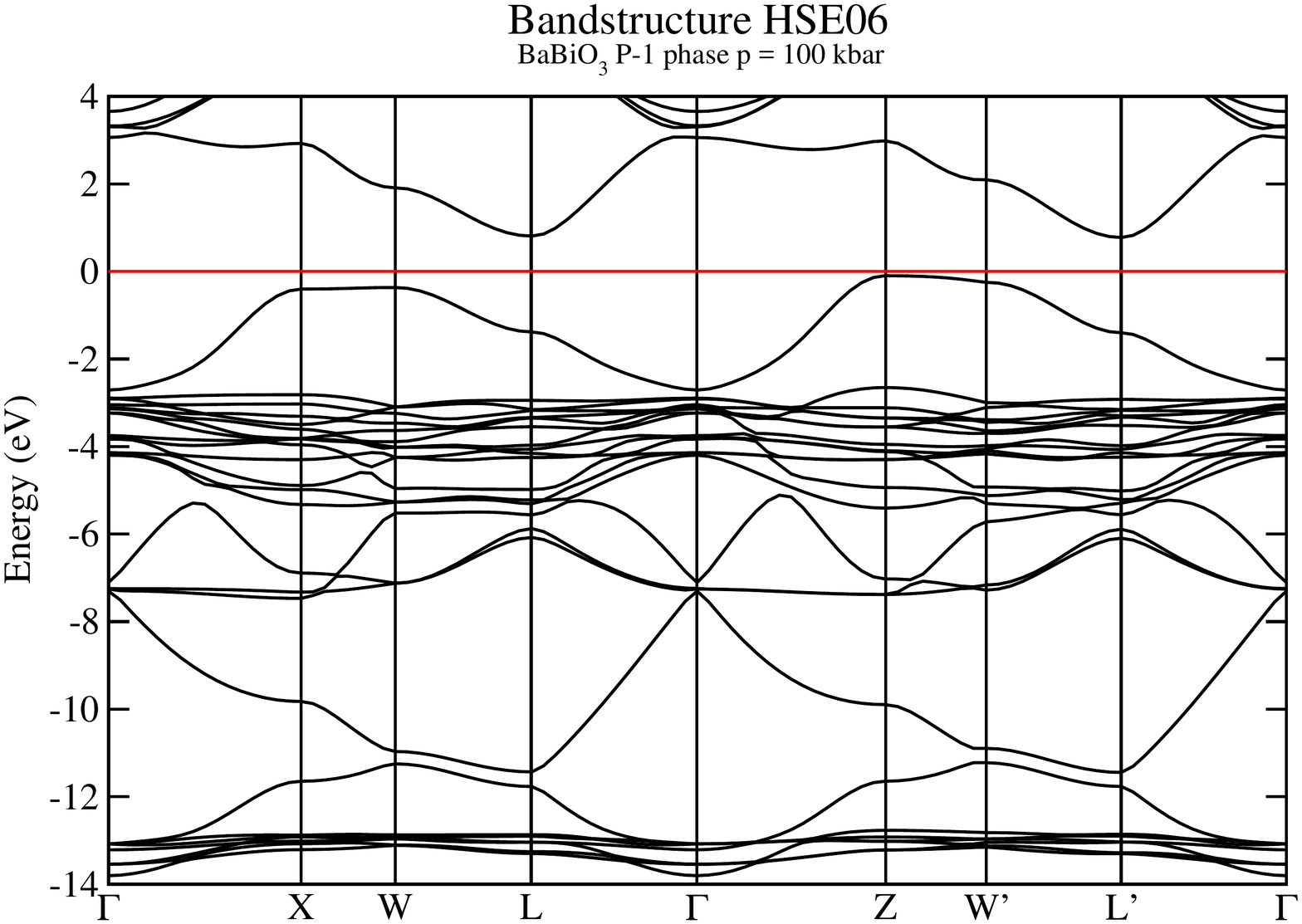}
    \includegraphics[width=0.45\textwidth]{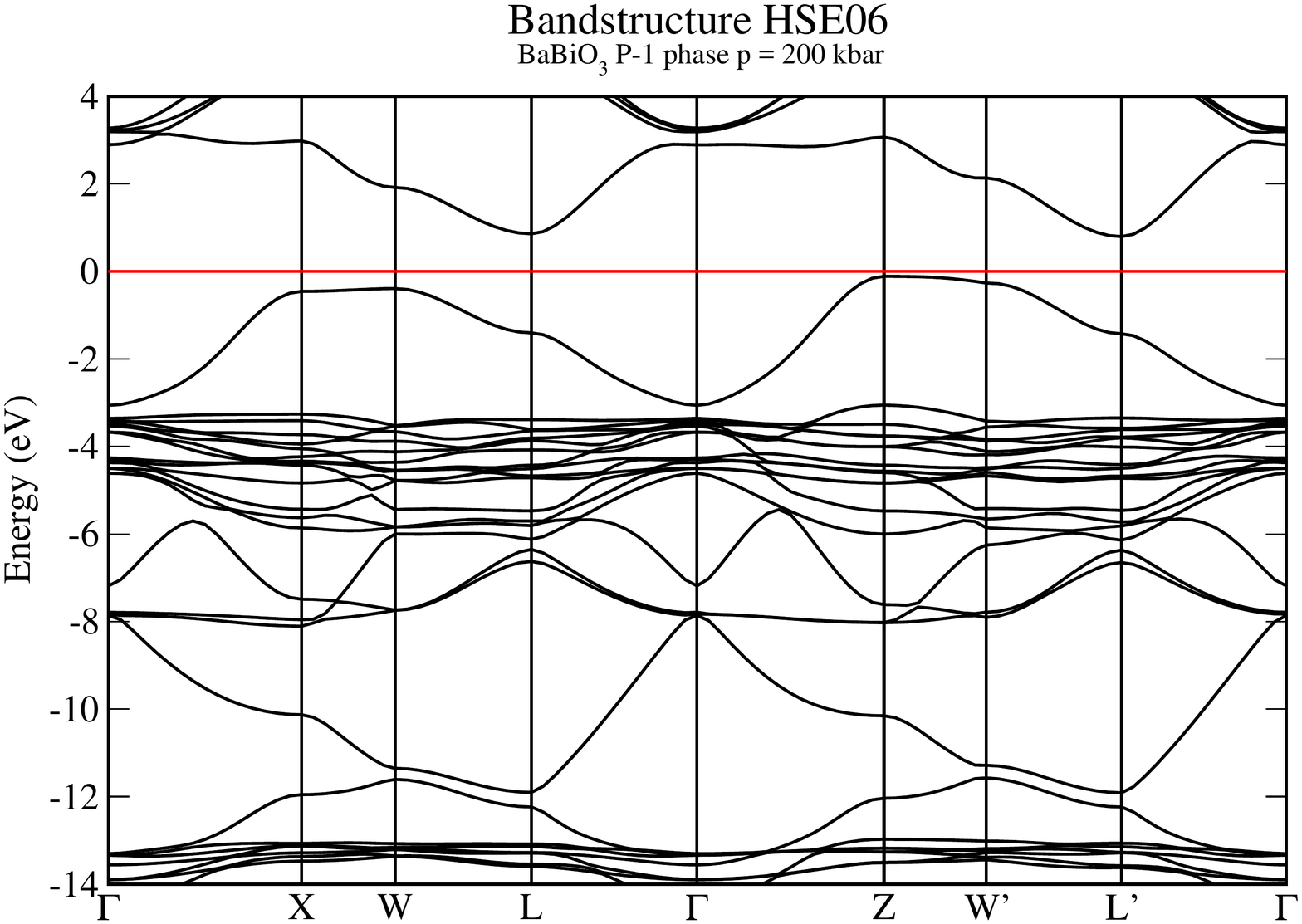}
    \includegraphics[width=0.2\textwidth]{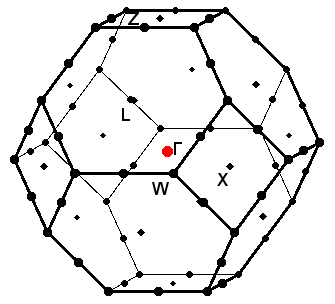}
    \caption{HSE electronic structure of \textit{P-1} BaBiO$_3$
at $P$= 10 and 20 GPa. On the bottom panel the Brillouin zone is shown.}
    \label{fig:bands_P-1}
  \end{figure}

As anticipated, in none of the phases theoretically predicted for high
pressure BaBiO$_3$ all the Bi ions become equivalent. In line with
that result, we expect that all these phases should be insulating, or
nearly insulating.  In order to check this important aspect we
performed the electronic band structure calculation using the hybrid
functional HSE06 -- the method of choice which gave correct results at
low pressures.  Since both \textit{I2/m} and \textit{P-1} phases
represent relatively small distortions of the ideal perovskite
structure the Brillouin zone of both phases is similar to that of fcc
phase (Fig.\ref{fig:bands_P-1}, see also Fig.1 in
Ref.\cite{Mattheiss1983}).  We calculated the band structure of both
phases and found it to be qualitatively similar to that found for
artificially disproportionated ideal perovskite structure without
octahedra rotation by Thonhauser and Rabe\cite{Thonhauser2006}. In
their idealized case the indirect band gap in the LDA approximation
opened up between the W and L points. In our case where we use the
real energy-minimized HSE06 \textit{I2/m} structure, the highest
occupied band between the W and X points is remarkably flat. This
implies that the indirect band gap is likely to open between the
valence band maximum (VBM) in the segment between W and X and
conduction band minimum in L. Due to symmetry breaking related to the
rotation of octahedra, however, degeneracy of bands at special points
in BZ in ideal cubic perovskite structure is lifted, as can be seen in
Figs.\ref{fig:bands_I2/m} and \ref{fig:bands_P-1}. In the
\textit{I2/m} structure the VBM is at the point X while in the
\textit{P-1} phase it is at the point Z on the pseudotetragonal axis.

\begin{figure}[ht]
    \centering
    \includegraphics[width=0.45\textwidth]{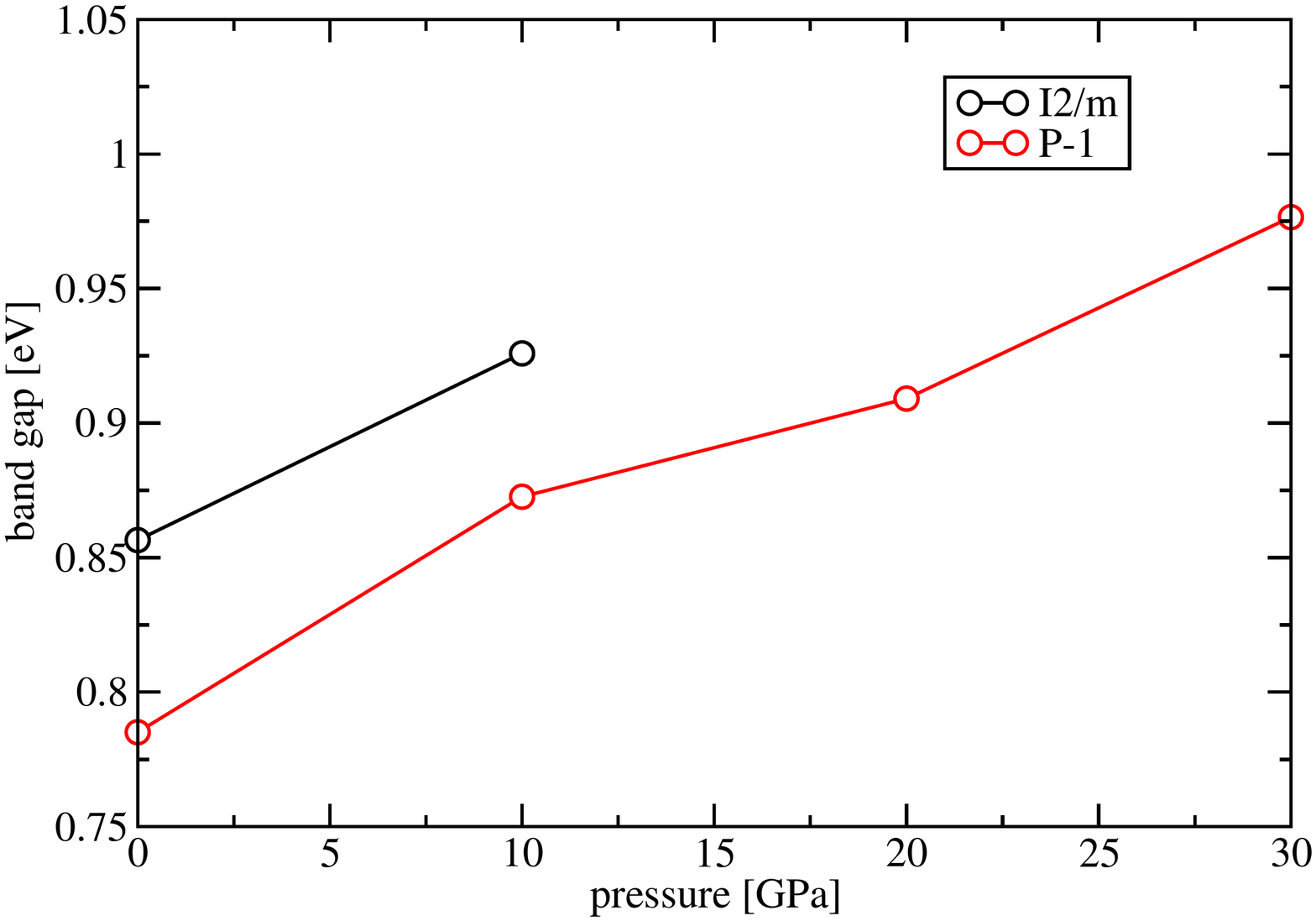}
    \caption{Pressure dependence of HSE indirect band gap of BaBiO$_3$ in \textit{I2/m} and \textit{P-1} phases.}
    \label{fig:gap}
  \end{figure}

The pressure dependence obtained for HSE06 equilibrium electronic
structures is shown for \textit{I2/m} at $P$=0 and 10 GPa in 
Fig.\ref{fig:bands_I2/m} and for \textit{P-1} at $P$=10 and 20 GPa in 
Fig.\ref{fig:bands_P-1}. The evolution is mild, and, as anticipated, there 
is no metallization under pressure. Despite the pressure-induced decrease
of the relative bond disproportionation, there is in fact a slight
increase of the electronic band gap, as shown in
Fig.\ref{fig:gap}. While that for \textit{I2/m} does not explain the
data below 10 GPa the theoretical gap increase predicted for the
\textit{P-1} phase agrees well with the observed resistivity increase
between 10 and 40 GPa. Quantitatively, we predict a slope of about 52
meV/10 GPa which is rather satisfactory compared to experimental slope
of 30 meV/10 GPa if one considers that even HSE06 functional is not
designed to provide the exact value of the band gap. In summary,
theory and experiment agree on BaBiO$_3$ becoming even more insulating
upon pressure increase above 10 GPa. The experimental slight decrease
of resistivity below 10 GPa could be caused by thermal fluctuations
affecting the rotation angle of the octahedra and in turn the band gap
-- temperature effects, here neglected, are expected to be more
important precisely at low pressures.

%
  
The \textit{C2/m} phase at 40 and 50 GPa is semimetallic (valence and
conduction band overlap by about 0.11 eV) -- but as was said above the
mixed-structure character of this phase suggests that it is probably
small-cell artifact.  Mixed structures arising for this reason are not
uncommon in insulators, due to broken bonds at the interface.  The
\textit{P1} phase at 50 GPa has instead a large band gap of 2.95 eV
and is a good insulator.  If indeed the \textit{C2/m} phase is an
artifact and \textit{P-1} survives even if metastable to 40-50 GPa,
then at 50 GPa and beyond some disordered structures similar to
insulating \textit{P1} might take over.  The \textit{P1} gap at 40 GPa
it is 3.01 eV so it increases upon decreasing pressure, which agrees
with the slope of experimental resistivity upon decompression down to 10 GPa 
where the disordered system might start recrystallizing.

\section{Discussion and Conclusions}
\label{sec:discussion}

Our main result is that BaBiO$_3$ resists all tendencies towards
equivalent Bi(4+) ions under pressure. As pressure increases,
structure changes so as to preserve two (or at least two) inequivalent
valence states of the Bi ion, and either a frank insulating or a
nearly insulating character along with that.  This outcome may be seen
as a surprise if the disproportionation of BaBiO$_3$ was interpreted
as a Peierls-type CDW, suggesting that the persistent inequivalence of
two Bi sites should be attributed to a different origin. One possible
origin, which has been raised before, is an intrinsic intra-atomic
tendency of the Bi ion to avoid valency 4+, favoring either 3+ or 5+.

Further progress should be first of all experimental, identifying and
confirming the high pressure phases, {\it in primis} our predicted
pseudotetragonal \textit{P-1} phase (or the closely related tetragonal
\textit{I4/m} phase) between 7 and 27 GPa and possibly beyond till 40
GPa by high-quality X-ray or neutron diffraction.  Above 40 GPa, that
search may be considerably complicated by the high levels of
hysteresis which we found in resistivity, indicating that transitions
to higher pressure structures might require a non-trivial
rearrangement of atoms and reconstruction of structure.  In closing,
we can connect with a purely theoretical parallel study by another
group~\cite {Smolyanyuk2017} where a similar sequence of phases has
also been predicted, and which agrees with our overall conclusion of
inequivalent Bi ions and (with the exception of \textit{C2/m}, that
does not show up in our experiments) that there is no metallic state
in high-pressure BaBiO$_3$.

\begin{acknowledgments}
This work was supported by the Slovak Research and Development Agency
under Contract No.~APVV-15-0496, by the VEGA project No.~1/0904/15 and
by the project implementation 26220220004 within the Research \&
Development Operational Programme funded by the ERDF. Part of the
calculations were performed in the Computing Centre of the Slovak
Academy of Sciences using the supercomputing infrastructure acquired
in project ITMS 26230120002 and 26210120002 (Slovak infrastructure for
high-performance computing) supported by the Research \& Development
Operational Programme funded by the ERDF. Work in Trieste was partly
sponsored under ERC MODPHYSFRICT Contract 320796. We are grateful to
Dr. Lilia Boeri for sharing with us the unpublished results of
Ref. \cite {Smolyanyuk2017}.

\end{acknowledgments}

\end{document}